\documentclass[aps,amssymb,10pt,showpacs,showkeys,letterpaper]{revtex4}
\usepackage{graphicx}
\usepackage{amsmath}
\usepackage{epsfig}
\usepackage{bm}

\newcommand{\be}{\begin{equation}}
\newcommand{\ee}{\end{equation}}
\newcommand{\beq}{\begin{eqnarray}}
\newcommand{\eeq}{\end{eqnarray}}

\begin{document}

\title{Motion of charged particles on the Reissner-Nordstr\"{o}m (Anti)-de Sitter black holes}

\author{ Marco Olivares }
 \email{marco.olivares@usach.cl}
\affiliation{\it Departamento de F\'{\i}sica, Facultad de Ciencia,
Universidad de Santiago de Chile,\\ Casilla 307, Santiago 2, Chile}

\author{ Joel Saavedra }
 \email{joel.saavedra@ucv.cl}

\affiliation{\it Instituto de F\'{\i}sica, Pontificia Universidad
de Cat\'{o}lica de Valpara\'{\i}so,\\ Av. Universidad 330, Curauma,
Valpara\'{\i}so , Chile}

\author{ J.R. Villanueva }
 \email{jrvillanueval@uta.cl}
\author{ Carlos Leiva }
\email{cleivas@uta.cl}
\affiliation{\it Departamento de F\'{\i}sica, Facultad de Ciencias, Universidad de Tarapac\'{a},\\
 Av. General Vel\'{a}squez 1775, Arica, Chile}

\date{\today}

\begin{abstract}
In this work we address the study of movement of charged particles
in the background of charged black holes with non-trivial
asymptotic behavior. We compute the exact trajectories for
massive-charged particles in terms of elliptic Jacobi functions.
Finally we obtain a detailed description of orbits for
Reissner-Nordstr\"{o}m (Anti) de Sitter black holes in terms
of charge, mass and energy of the particles.
\end{abstract}



\maketitle

\section{Introduction}

Motion of particles on black holes is one interesting phenomena of
classical gravity. At this respect, there are several studies about
geodesic motion in the vicinity of black holes. For instance, the
geodesic structure of the Schwarzschild (S), Reissner-Nordstr\"{o}m (RN)
and Kerr (K) black holes were studied in detail by Chandrasekhar
\cite{chandra}. There, the author studied  motion of the test
particles  using  the Lagrangian and the Newmann-Penrose
formalism. If we think this problem in a modern point of view, it
is of high interest to study the motion of particles in the
vicinity of black holes with  asymptotic behaviors others than the
Schwarzschild case (flat case). Then, the so called SAdS or SdS black
holes have been objects of highly consideration, due to the
AdS-CFT conjecture, where there are  powerful tools that relate
gravitational theories with an asymptotically AdS behavior with a
conformal field theories of one less dimension \cite{maldacena}.
Besides, this kind of black holes provides a theoretical laboratory
for understanding a lot of important points in black holes physics and
 gravity theories and their extensions. %[lista de referencias].
 It is possible to generalize the asymptotic flat  spacetimes by
including a non-zero cosmological constant term. For instance, the
Kotler solution \cite{kotler} is obtained by imposing $R_{\mu
\nu}=\Lambda g_{\mu \nu}$ in the field equations. Some important implications of those can be found
in \cite{Nakao,Hledik2}. Different focuses
in the research can be developed depending on the value of the cosmological
constant: the Schwarzschild de Sitter case (SdS) (which is
obtained by considering a positive value of the cosmological constant  $\Lambda
> 0$), or the Schwarzschild anti-de Sitter case (SAdS) (which is
obtained by considering a negative value of the cosmological constant,
consider, $\Lambda < 0$). There are some previous works about the
geodesic motion in the SdS case \cite{Jaklitsch,podolsky,Stuchlik}. On
the other hand, when a negative cosmological constant, $\Lambda <
0$, is taking in to account i.e. Schwarzschild Anti de Sitter case
(SAdS), the studies are more complicated. However different aspects of
the geodesic structure of this spacetime can be found in
\cite{K-W,Kraniotis,COV}. The motion of (neutral) particles in RN black hole with non-zero cosmological constant has been
studied in \cite{Hledik}. Furthermore, in Ref. \cite{Hackmann:2008tu,
Hackmann:2008zz, Hackmann:2008zza}, authors presented  analytical
solutions of the geodesic equation of massive test particles in
higher dimensional Schwarzschild, Schwarzschild (anti) de Sitter ,S(A)dS,
Reissner Nordstr\"{o}m, RN, and Reissner Nordstr\"{o}m (anti) de Sitter, RN(A)dS
spacetimes and they obtained  complete solutions  and a classification
of the possible orbits in these geometries in term of
Weierstrass functions.
Also, the ecuatorial circular motion in Kerr- de Sitter spacetime is studied in \cite{Slany}.

In this article we are interested in the study of the motion of charged
particles on the Reissner-Nordstr$\ddot{o}$m (Anti) de Sitter
black hole. In doing so, we start  considering the
Hamilton-Jacobi formalism in order to write equations of motion
and then we reduce our problem to quadratures. Then we solve
quadratures equation and we found all the possible orbits for the
geometries under consideration in terms of the elliptic Jacobi
functions.

\section{Motion of Charged Particles in the Vicinity of Black Hole }

We are interested in the study of the motion of massive-charged particles near
to an static, spherically symmetric and charged black hole when a non-zero
cosmological constant is taken into account. The $\Lambda > 0$ case is called
the Reissner-Nordstr\"{o}m de Sitter spacetime (RNdS), while the $\Lambda < 0$ case is call it
the Reissner-Nordstr\"{o}m Anti de Sitter spacetime (RNAdS).

In terms of the usual Schwarzschild coordinates ($t, r, \theta, \phi$),
the metric is written as

\begin{equation}
ds^{2}=-f(r)dt^{2}+\frac{dr^{2}}{f(r)}+r^{2}(d\theta^{2}+sin^{2}\theta
d\phi^{2}), \label{i.1} \end{equation}

\noindent where  $f(r)$ is the lapsus function given by

\begin{equation} f(r)=1-\frac{2M}{r}+\frac{Q^{2}}{r^{2}}-\frac{\Lambda
r^{2}}{3},\label{i.2}\end{equation}

\noindent and the coordinates satisfy the relations: $-\infty \leq
t \leq \infty$, $r\geq0$, $0\leq\theta\leq\pi$ y  $0\leq\phi\leq
2\pi$.

 %%%%%%%%%%%%%%%%%%%%%%%%%%%%%%%%%%%%%%%%%%%%%%%%%%%%%%%%%%%%%%%%%%

We would like to consider the motion of test particles with mass $m$
and charge $q$ in the framework of general relativity. In order to
obtain the equation of motion, we apply the Hamilton-Jacobi formalism. In
this sense, the Hamilton-Jacobi equation  for the geometry described by the metric $g_{\mu\nu}$ is
\begin{equation} g^{\mu \nu}\left(\frac{\partial S}{\partial
x^{\mu}}+qA_{\mu}\right)\left(\frac{\partial S}{\partial
x^{\nu}}+qA_{\nu}\right)+m^{2}=0,\label{i.3}\end{equation}

where $A_{\mu}$ represents the vector potential components associated with
the charge of the black hole (because, we are considering charged
static black holes where the only non-vanishing component of the vector
potential is the temporal $A_{0}=\frac{Q}{r}$),  $S$
corresponds to the characteristic Hamilton function. Considering our
metric this equation can be written as

\begin{equation} -\frac{1}{f(r)}\left(\frac{\partial S}{\partial t}+\frac{q
Q}{r}\right)^{2}+f(r)\left(\frac{\partial S}{\partial
r}\right)^{2}+\frac{1}{r^{2}}\left(\frac{\partial S}{\partial
\theta}\right)^{2}+\frac{1}{r^{2}sin^{2}\theta}\left(\frac{\partial
S}{\partial \phi}\right)^{2}+m^{2}=0,\label{i.4}\end{equation}

\noindent in order to solve this equation we use the following ansatz
\begin{equation} S=-E t+S_{1}(r)+S_{2}(\theta)+J \phi,\label{i.5}\end{equation}
where $E$ and $J$ are identified as the energy and angular momentum
of the particle. Using this ansatz  Eq.(\ref{i.4}) reads as follow
\begin{equation} -\frac{r^{2}}{f(r)}\left(-E+\frac{q
Q}{r}\right)^{2}+r^{2}f(r)\left(\frac{\partial S_{1}}{\partial
r}\right)^{2}+\left(\frac{\partial S_{2}}{\partial
\theta}\right)^{2}+J^{2}csc^{2}\theta+r^{2}m^{2}=0.\label{i.6}\end{equation}

\begin{figure}[!h]
 \begin{center}
   \includegraphics[width=150mm]{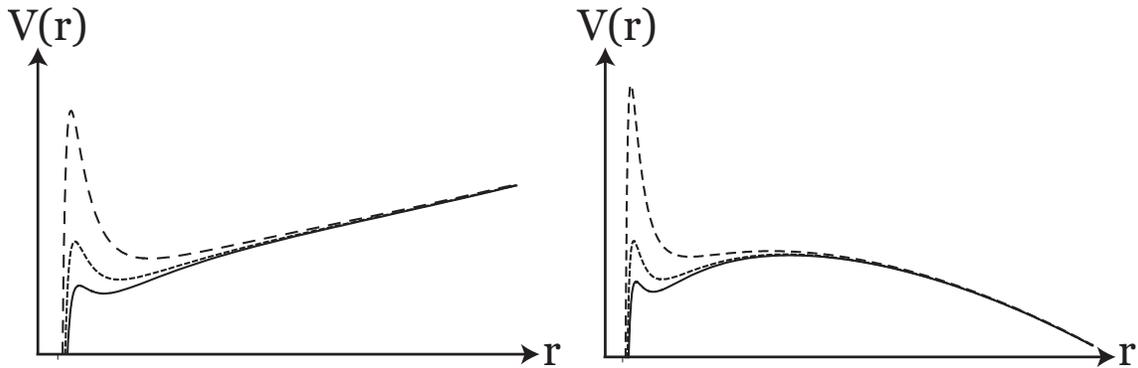}
 \end{center}
 \caption{Effective potential for charged particles on the geometry of RN for different values of angular momentum, $L$ (left panel SAdS case and right panel SdS case). Both of them were plotted considering $Q=0.85M$, $m=0.2M$, $q=0.18M$ and $\ell= 400M$. }
 \label{fig.1}
\end{figure}

Using the standard procedure we recognize the following constant
\begin{equation} L^{2}=\left(\frac{\partial S_{2}}{\partial
\theta}\right)^{2}+J^{2}csc^{2}\theta,\label{i.7}\end{equation}
without lack of generality we consider
that the motion is developed in the invariant plane
$\theta=\frac{\pi}{2}$ and in this case $L$ is equal to angular
momentum $J$. Then, we obtain the equation of motion
\begin{equation} -\frac{r^{2}}{f(r)}\left(-E+\frac{q
Q}{r}\right)^{2}+r^{2}f(r)\left(\frac{\partial S_{1}}{\partial
r}\right)^{2}+L^{2}+r^{2}m^{2}=0,\label{i.8}\end{equation}
and thus, we find  formal solutions for the radial
component of the action
\begin{equation} S_{1}(r)= \epsilon \int \frac{dr}{f(r)} \sqrt{\left(E-\frac{q
Q}{r}\right)^{2}-f(r)\left(m^2+\frac{L^{2}}{r^{2}}\right)},\label{i.9}\end{equation} where $\epsilon=\pm 1$.
Now, using the Hamilton-Jacobi method, we simplify our
study to the following quadrature problem
\begin{equation} t=\epsilon \int \frac{dr}{f(r)} \left(E-\frac{q
Q}{r}\right)\left[\left(E-\frac{q
Q}{r}\right)^{2}-f(r)\left(m^2+\frac{L^{2}}{r^{2}}\right)\right]^{-1/2},\label{i.10}\end{equation}
where we can obtain the radial velocity to respect the coordinate
time
\begin{equation} \frac{dr}{dt}=\pm  \frac{f(r)}{\left(E-\frac{q Q}{r}\right)}
\sqrt{\left(E-\frac{q
Q}{r}\right)^{2}-f(r)\left(m^2+\frac{L^{2}}{r^{2}}\right)}.\label{i.11}\end{equation}
Now,  the condition of  turning point $(\frac{dr}{dt})_{r=r_t}=0$
allows us to define an effective potential. In fact, considering
that

\begin{equation}
\frac{f(r)}{\left(E-\frac{q Q}{r}\right)}\neq 0, \forall r,
\end{equation}

we can factorize the term under the square root as:

\begin{equation}
\left(E-\frac{q
Q}{r}\right)^{2}-f(r)\left(m^2+\frac{L^{2}}{r^{2}}\right)=(E-V_-)(E-V_+),\label{i.12}\end{equation}

where we can recognize the effective potential for the particle
with mass $m$ and electric charge $q$ as

\begin{equation} V_{\pm}(r)=\frac{q
Q}{r}\pm\sqrt{f(r)}\sqrt{m^{2}+\frac{L^{2}}{r^{2}}},\label{i.13}\end{equation}
therefore, eq.(\ref{i.11}) can be written as

\begin{equation} \frac{dr}{dt}=\pm  \frac{f(r)}{\left(E-\frac{q Q}{r}\right)}
\sqrt{\left[E -
V_{-}(r)\right]\left[E-V_{+}(r)\right]}.\label{i.14}\end{equation}
Since the charge of the test particle is much smaller than the
mass ($q\ll M$), eventually $V_{-} <0$ and for this reason we
choose the positive branch of the effective potential:
$V_{eff}=V_{+}\equiv V$. The behavior of effective potential is
shown in Fig.\ref{fig.1} for RNdS and RNAdS cases with different
values of the angular momentum, but kepping $m, q$ fixed. It's
worth noting that the particle still has electric potential
energy at the horizon.

Finally, we can classify different kinds of motion for
massive-charged particles through the values of the  $L$ in the
following way

\begin{itemize}
    \item  Motion of charged particles with angular momentum
    \item  Motion of charged particles with vanishing angular momentum
\end{itemize}

In the following sections we study in detail the orbits before
mentioned.

 \subsection{ Circular Orbits }
The orbits can be classified by their values of energy and angular
momentum. In order to have at least a stationary system  the effective potential $V(r)$ have to exhibit  extremes
for fixed values of radial coordinate, $r=r_x$,
\begin{equation} \frac{d V(r)}{dr}|_{r=r_x}=0.\label{C.2}
\end{equation}

For simplicity, we rewrite the effective potential as

\begin{equation} V(r)= h(r) + \sqrt{f(r)\, j(r)}, \label{C.21}\end{equation}

\noindent where $f(r)$ is the lapsus function, and

$$
h(r)=\frac{q Q}{r},\quad j(r)= m^{2} + \frac{L^{2}}{r^{2}}.
$$

Therefore, using eq. (\ref{C.21}) into eq. (\ref{C.2}) yields

\begin{equation} \left[h'(r)\sqrt{f(r)\, j(r)}+ \frac{1}{2}(f'(r)\,j(r) + f(r)\,j'(r))\right]_{r=r_x}=0 \label{C.22}\end{equation}

Notice that this equation leads to a polynomial of twelfth grade, so, their solution is restricted to the numeric plane.
However, it is possible to determine the periods of revolution of the circular orbits, both stable and unstable, with respect to the proper time, $\tau$, and coordinate time, $t$, in the following way:  the condition (\ref{C.22}) allows to obtain the angular momentum for the stable and unstable circle orbits, $L_x=L_s$ and $L_x=L_c$ (at $r_x=r_s$ and $r_x=r_c$
in FIG. \ref{fig.2}, respectively; and $r_x=r_e$ and $r_x=r_i, R_i$ in FIG. \ref{fig7}, respectively). In our case, it is given by

 \begin{equation} a+b\,L_x^{2}=[c+d\,L_x^{2}]^{2},\label{C.23}\end{equation}

\noindent where

\begin{equation}  a= m^{2}f(r_{x})\,h'(r_{x})^{2};\qquad b= \frac{f(r_{x})\,h'(r_{x})^{2}}{r^{2}}; \qquad c=
\frac{m^{2}f'(r_{x})}{2};\qquad d=
\frac{f'(r_{x})}{2r^{2}}-\frac{f(r_{x})}{r^{3}}\label{T.24}\end{equation}

 Thus, the real solution of the quadratic equation for $L_x$ give us the angular momentum of the circular orbit. Explicitly, the constants of motion, $L_x$ and $E_x$, for the circular orbits are given by

\begin{equation} L_x= \left[\frac{b-2cd-\sqrt{b^{2}+4d(da-bc)}}{2d^{2}}\right]^{1/2}, \label{T.25}\end{equation}

\noindent and

\begin{equation} E_x = h(r_x) + \sqrt{f(r_x)}\left[m^{2}+\frac{b-2cd-\sqrt{b^{2}+4d(da-bc)}}{2d^{2}r^{2}}\right]. \label{T.26}\end{equation}

Therefore, the proper period of the circular orbit ($T_{\tau}=\frac{2\pi r_{x}^{2}}{L_x}$) is

\begin{equation} T_{\tau}=2\pi r_{x}^{2}\left[\frac{b-2cd-\sqrt{b^{2}+4d(da-bc)}}{2d^{2}}\right]^{-1/2},\label{T.27}\end{equation}

\noindent and the coordinate period ($T_{t}= \frac{2\pi r_{x}^{2}}{L_x f(r_x)}(E_x-h(r_x))= T_{\tau} \sqrt{\frac{j(r_x)}{f(r_x)}}$) is

\begin{equation} T_{t}= \frac{T_{\tau}}{\sqrt{f(r_x)}} \left[m^{2}+\frac{b-2cd-\sqrt{b^{2}+4d(da-bc)}}{2d^{2}r_{x}^{2}}\right]^{1/2}. \label{T.28}\end{equation}

Notice that, if neutral particles are taken into account (i. e. $q=0$ and $h(r_x) = 0$), the proper and coordinate period are given by

\begin{equation} T_{\tau}=\frac{2\pi r_{x}}{m}\sqrt{\frac{r_{x}^{2}-3Mr_{x}+2Q^{2}}{Mr_{x} - Q^{2}- \Lambda r_{x}^{4}/3}},\label{T.29}\end{equation}

\noindent and

\begin{equation} T_{t} = \frac{T_{\tau}}{\sqrt{f(r_x)}}\left[\frac{r_{x}^{2}- 2Mr_x + Q^{2}-\Lambda r_{x}^{4}/3 }{r_{x}^{2}-3Mr_{x}+2Q^{2}}\right]^{1/2},  \label{T.30}\end{equation}

respectively. Recently, the authors in \cite{pugliese} have studied other aspect of the circular motion of the neutral particles
in the RN spacetime such as the stability of the orbits.

%%%%%%%%%%%%%%%%%%%%%%%%%%%%%%%%%%%%%%%%%%%%%%%%%%%%%%%%%%%%%%%%%%%%%%%%%%%%%%%%%%%%%%5
\section{ Charged Particles on the geometry of  Reissner-Nordstr\"{o}m  Anti-de Sitter black hole}

First at all, let's focus in the motion of charged particles on
the geometry of  Reissner-Nordstr\"{o}m  Anti-de Sitter black
hole. In this case charged particles are affected by an effective
potential described in right panel of Fig. 1. Considering motion
with $L\neq 0$ we obtain the equation
\begin{equation}
\dot{r}^{2}=[E - V_{-}(r)][E-V(r)]=-\left(\frac{m}{\ell}\right)^{2}\frac{P_6(r)}{r^{4}},\label{ads.21}\end{equation}
where
\begin{equation}
P_{6}(r)=\sum_{j=0}^{6}p_j r^{j},\label{ads.22}
\end{equation}
and the coefficients are
$$
p_0=\frac{Q^{2}L^{2}\ell^{2}}{m^{2}},\quad p_1=-\frac{2ML^{2}\ell^{2}}{m^{2}},\quad
p_2=\ell^{2}\left(\frac{L^{2}}{m^{2}}+Q^{2}-\frac{q^{2}Q^{2}}{m^{2}}\right),
$$
$$
p_3=-2\ell^{2}\left(M-\frac{qQE}{m^{2}}\right),\quad
p_4=-\ell^{2}\left(\frac{E^{2}}{m^{2}}-1-\frac{L^{2}}{m^{2}\ell^{2}}\right),\quad
p_5=0,\quad p_6=1.
$$

\begin{figure}[!h]
 \begin{center}
   \includegraphics[width=100mm]{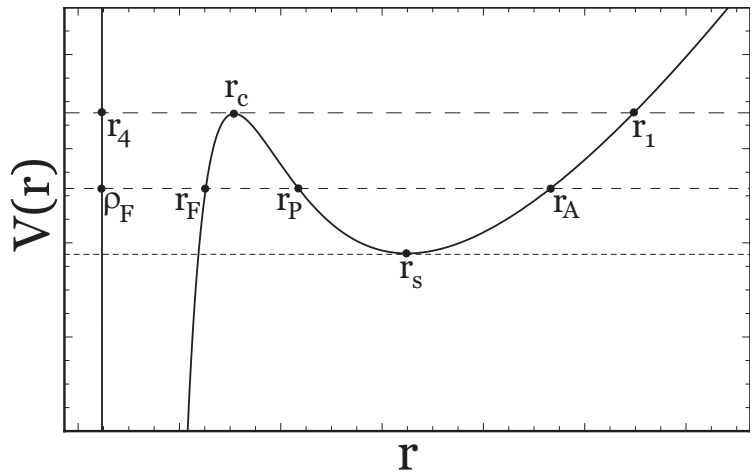}
 \end{center}
 \caption{Effective potential for charged particles on the geometry of RNAdS for $M=1$, $Q=0.85M$, $\ell=400M$, $L=0.33M^{2}$, $q=0.18M$ and $m=0.2M$. }
 \label{fig.2}
\end{figure}

The  solutions (different and reals) of this sixth degree
polynomial, correspond to the physical distance that characterize
the motion for  the so called periastron, apastron and circular
orbit radii. Physical reality of orbits depends on the values of
constants $E,M,Q,q$ and $L.$ We can distinguish in our description
two sets of values for the constants in order to classify possible
motions. One of them allows the existence of planetary orbits, in
this case the polynomial has six real distances. Besides, this set
has a second class trajectory that represents free fall to the
event horizon. The second set of fixed values for the constants
correspond to critical orbits, for example, here  we have the
unstable circle orbit. Finally, we discuss five trajectories
that have physical meaning for the geometry under consideration
(see FIG. \ref{fig.2} for a better visualization of the orbits.)
\begin{itemize}
 \item  Planetary Orbit: In this case the orbit corresponds to a bounded trajectory  that exhibit oscillation between two extremal distances:
 the {\it periastron} and the {\it apastron}, ($r_P$ and $r_A$ in FIG. \ref{fig.2}, respectively). For simplicity, we shall consider the case of a double-degenerate real negative root, $\sigma_F$, such that
 the polynomial  $P_6$ can be writen as $P_6(r)=(r-\sigma_F)^{2} P_4(r)$. This case has also a
particular solution called Reissner-Nordstr\"{o}m limit.

 \item  Second Kind Trajectory: This trajectory
 is computed with the same parameter than the planetary orbits. It
 corresponds to a trajectory that starts  at rest from a
 finite distance, $r_F$. This kind of motion represents the fall to the event
horizon, and it is considered that has a turning point inside the
Cauchy horizon, $\rho_F$.

\item  Critical Trajectory: There are trajectories of the first and
second kind with angular momentum  $L_c$, given by (\ref{co.17}),
and energy $E_c$ given taken $r=r_c$ in  (\ref{i.12}) and equating
it to zero. The first one starts at rest at a finite distance
outside from the unstable circular orbit, $r_1$, and then it approximates
asymptotically to it. The second kind  approximates to the
unstable orbit from the inside of it.

 \item  Radial Trajectory: These trajectories have null angular momentum and physically they describe
radial fall from rest to the event horizon.

\item  Circle Orbits: For some fixed values of the constants it is possible to find solutions of the equation of motions that represent stable
  and unstable circle orbits. This case was discussed in the previous section and it is possible to find the connection with the Lyapunov exponent
  for the unstable orbit with the quasinormal modes [cardoso, work in progress].

\end{itemize}

 \subsection{Planetary Orbit}
\begin{figure}[!h]
 \begin{center}
   \includegraphics[width=160mm]{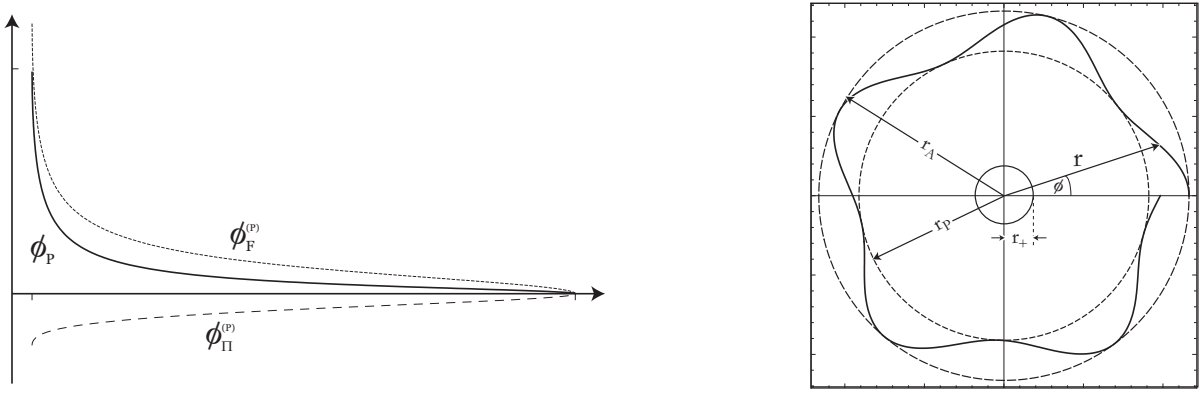}
 \end{center}
 \caption{Polar plot for the planetary orbit of the test particle in the Reissner-Nordstr\"{o}m Limit. Clearly,
 there is a precession of the perihelion, which is given explicitly by eq.(\ref{ads.32}). }
 \label{fig.3}
\end{figure}

 In order to characterize this motion we rewrite the polynomial
 $P_4(r)$ in term of his roots
\begin{equation}
P_4(r)=(r-\rho_{S})(r-r_{S})(r-r_{P})(r_{A}-r).\label{ads.23}\end{equation}
where $ \sigma_{F}<0<\rho_{S}<r_{S}<r_{P}<r_{A}$, and we identify  periastron and apastron distances as $r_{P}$ and
$r_{A}$ respectively. These  trajectories are defined  inside the following limits
$ r_{P}<r<r_{A}$, and the equation of
motion  has the following quadrature
\begin{equation} - \frac{dr}{d \phi}=\frac{\sqrt{P_6(r)}}{ \ell
L_{P}},\label{ads.24}\end{equation}

\noindent where $L_P$ is the angular momentum of the test particle in the planetary orbit.
The corresponding integral is given by
\begin{equation} \phi(r)=\ell
L_{P}\int^{r}_{r_{A}}\frac{-dr}{(r-\sigma_{F})\sqrt{P_4(r)}},\label{ads.25}\end{equation}
now introducing  constants $\alpha_{P}=\frac{2\ell
L_{P}}{\sqrt{(r_{A}-r_{S})(r_{P}-\rho_{S})}}$,
$\beta_{P}=\frac{1}{(\rho_{S}-\sigma_{F})}$ y $\gamma_{P}=-\frac{
(r_{A}-\rho_{S})}{(r_{A}-\sigma_{F})(\rho_{S}-\sigma_{F})}$, it is
possible to get  the solution  given by a Jacobi elliptic integrals
of first and third kinds
\begin{equation} \phi_P(r)=\phi_{F}^{(P)}+\phi_{\Pi}^{(P)},\label{ads.26}\end{equation}
where $\phi_{F}^{(P)}=\alpha_{P}\beta_{P} F(\psi_{P};\kappa_{P})$ and $\phi_{\Pi}^{(P)}= \alpha_{P}\gamma_{P}\Pi
(\psi_{P};\kappa_{P},n_{P})$, with the elliptic parameters given by
\begin{equation} \psi_{P}=\arcsin
\sqrt{\frac{(r_{P}-\rho_{S})(r_{A}-r)}{(r_{A}-r_{P})(r-\rho_{S})}},\label{ads.27}\end{equation}

\begin{equation}
\kappa_{P}=\sqrt{\frac{(r_{A}-r_{P})(r_{S}-\rho_{S})}{(r_{A}-r_{S})(r_{P}-\rho_{S})}},\label{ads.28}\end{equation}

\begin{equation}
n_{P}=\frac{(r_{P}-r_{A})(\sigma_{F}-\rho_{S})}{(r_{P}-\rho_{S})(\sigma_{F}-r_{A})},\label{ads.29}\end{equation}

similar results can be found in \cite{Hackmann:2008tu,
Hackmann:2008zz, Hackmann:2008zza}, where authors presented analytical
solutions for the geodesic equation of massive test particles in
higher dimensional Schwarzschild, Schwarzschild–(anti)de Sitter,
Reissner–Nordstr¨om and Reissner–Nordstr¨om–(anti)de Sitter
space–times. In  FIG.\ref{fig.3} we show  orbits for an specific value of the angular
momentum and present the case of an elliptic orbit that
precesses between the periastron and apastron.

At this point, we can  calculate the precession of the
perihelion as
\begin{equation}
2 \phi(r_{P})=2\pi+\Delta\varphi ,\label{ads.30}\end{equation}

\noindent and considering the exact solution (\ref{ads.26}) we
obtain

\begin{equation}
\Delta\varphi=2\left[\phi_{\Pi}^{(P)}(r_P)+\phi_{F}^{(P)}(r_P)\right]-2\pi
,\label{ads.31}\end{equation}

One particular solution is obtained, when we perform a fine tuning in
the physical distance under consideration. This case corresponds
to the  called Reissner-Nordstr\"{o}m limit, that we are going to
discuss in the next subsection.

\subsubsection{Reissner-Nordstr\"{o}m Limit}
One approximated solution of the first order represents the limit
case of the motion on  Reissner-Nordstr\"{o}m black hole. This
limit corresponds to the solution when $\ell\rightarrow\infty$ and
therefore we must  consider the potential
$$
V(r)=\frac{q Q}{r}+\sqrt{\left(1-\frac{2M}{r}+\frac{Q^{2}}{r^{2}}\right)\left(m^{2}+\frac{L^{2}}{r^{2}}\right)},
$$
in which case we obtain two kinds of orbits:
\begin{itemize}
\item confined orbits: they are obtained when the relation $E<m$ is satisfied.
\item non-confined orbits: they are obtained when the relation $E\geq m$ is satisfied.
\end{itemize}
i).- \underline{Confined orbits}: defining $\zeta^2=(m^2 -
E^2)^{-1}$, the motion equation (\ref{ads.24}) can be written as
$$
\left(\frac{dr}{d\phi}\right)^{2}=
\frac{\textsc{P}_{4}(r)}{\zeta^{2}L^{2}},
$$
where,
$$
\textsc{P}_{4}(r)=\sum_{j=0}^{4} p'_{j}r^j
$$
and the coefficients are
$$
p'_0=-\zeta^2 Q^{2}L^{2};\quad p'_1=2\zeta^2 M L^{2};\quad p'_2=-\zeta^2(L^{2}+m^{2}Q^{2}-q^{2}Q^{2});\quad
p'_3=2\zeta^{2}(m^{2}M-qQE);\quad p'_4=-1,
$$
In terms of the roots of the polynomial,we can write
$$
p'_4=(r'_{A}-r)(r-r'_{P})(r-r'_{F})(r-\rho'_{F}).
$$
Then, the solution  can be written as
$$
\phi'_P(r)=\alpha'_{P} F(\psi'_{P};\kappa'_{P})
$$
where
$$
\alpha'_{P}=\frac{2 \zeta
L'}{(\sqrt{r'_{A}-r'_{F})(r'_{P}-\rho'_{F})}};
$$
\newline
$$
 \psi'_{P}=\arcsin
\sqrt{\frac{(r'_{P}-\rho'_{F})(r'_{A}-r)}{(r'_{A}-r'_{P})(r-\rho'_{F})}}
$$
\newline
$$
\kappa'_{P}=\sqrt{\frac{(r'_{A}-r'_{P})(r'_{F}-\rho'_{F})}{(r'_{A}-r'_{F})(r'_{P}-\rho'_{F})}}
$$

Defining the constants $\varpi=\frac{1}{\alpha'_{P}}$ and
$\delta=\frac{(r'_{A}-r'_{P})}{(r'_{P}-\rho'_{F})}$, we can write
the solution in terms of elliptic Jacobi sine
\begin{equation} r(\phi')=\frac{r'_{A}+\rho'_{F} \delta sn^{2}(\varpi
\phi')}{1+\delta sn^{2}(\varpi \phi')},\label{ads.31}\end{equation}

In our case, the precession of the perihelion (\ref{ads.30}) is given by

\begin{equation}
\Delta\varphi=2[\alpha'_{P}K(\kappa'_{P})-\pi]
,\label{ads.32}\end{equation}

\noindent where $K(\kappa'_{P})$ is the complete elliptic Jacobi integral of the first kind.

\subsection{Second Kind Trajectory}
This trajectory corresponds to the one where the test particle
starts at rest from a finite distance bigger than the event horizon
and then falls to it.  Its motion is developed in the following
region $\rho_{S}<r<r_{S}$. In this case
the corresponding angular quadrature is given by

\begin{equation} \phi(r)=\ell
L_{P}\int^{r}_{r_{S}}\frac{-dr}{(r-\sigma_{F})\sqrt{P(r)}},\label{ads.33}
\end{equation}

\noindent now we have a fourth degree polynomial described by

\begin{equation}
P(r)=(r-\rho_{S})(r_{S}-r)(r_{P}-r)(r_{A}-r).\label{ads.34}
\end{equation}

Using a similar procedure than the used for the planetary orbit,
 we  first define the following constants
 $\alpha_{s}=\frac{2\ell
L_{P}}{\sqrt{(r_{A}-r_{S})(r_{P}-\rho_{S})}}=\alpha_{P}$,
$\beta_{s}=\frac{1}{r_{P}-\sigma_{F}}$ and $\gamma_{s}=\frac{
r_{P}-r_{S}}{(\sigma_{F}-r_{P})(\sigma_{F}-r_{S})}$. The solution
can be written in terms of Jacobi elliptic functions as follows
\begin{equation} \phi_{s}(r)=\phi_F^{(s)}+\phi_{\Pi}^{(s)},\label{ads.35}
\end{equation}
\noindent where $\phi_F^{(s)}=\alpha_{s} \beta_{s} F(\psi_{s}; \kappa_{s})$ and $\phi_{\Pi}^{(s)}=\alpha_{s} \gamma_{s} \Pi(\psi_{s}; \kappa_{s}, n_{s})$, with

\begin{equation} \psi_{s}=\arcsin
\sqrt{\frac{(r_{P}-\rho_{S})(r_{S}-r)}{(r_{S}-\rho_{S})(r_{P}-r)}},\label{ads.36}
\end{equation}

\begin{equation}
\kappa_{s}=\sqrt{\frac{(r_{A}-r_{P})(r_{S}-\rho_{S})}{(r_{A}-r_{S})(r_{P}-\rho_{S})}}=
\kappa_{P},\label{ads.37}\end{equation}

\begin{equation}
n_{s}=\frac{(r_{S}-\rho_{S})(\sigma_{F}-r_{P})}{(r_{P}-\rho_{S})(\sigma_{F}-r_{S})},\label{ads.38}
\end{equation}

This kind of trajectory is shown in Fig.(4) where we show the
analytic continuation to the inner space to the event horizon. In
sum this trajectories are doomed to cross the event horizon.

\subsection{critical trajectory }
There are trajectories of the first and second kind. The first one
starts at rest at a finite distance outside from the unstable circular
orbit and then it approximates asymptotically to it. The second
kind  approximates to the unstable orbit from inner distance.

\subsubsection{First class critical trajectory}
Critical trajectory of the first kind corresponds to the motion of
particles to asymptotically tend to a circle orbit from a great
distance compared to the radius of this orbit. The region of this
motion is defined by the limits
$r_{c}<r<r_{1}$ and the respective polynomial
becomes
\begin{equation}
F_{c}(r)=(r-r_{6})(r-r_{5})(r-r_{4})(r-r_{c})^{2}(r_{1}-r).\label{ads.39}\end{equation}
then we can obtain the integral for the orbit as follows
\begin{equation} \phi_{c}^{(1)}(r)=\ell
L_{P}\int^{r}_{r_{1}}\frac{-dr}{(r-r_{c})\sqrt{(r_{1}-r)(r-r_{4})(r-r_{5})(r-r_{6})}}.\label{ads.40}\end{equation}
Defining the following constants $\alpha_{1}=\frac{2\ell
L_{P}}{\sqrt{(r_{1}-r_{5})(r_{4}-r_{6})}}$,
$\beta_{1}=\frac{1}{(r_{6}-r_{c})}$ and
$\gamma_{1}=\frac{(r_{6}-r_{1})}{(r_{1}-r_{c})(r_{6}-r_{c})}$ we
can obtain a solution for polar angle
\begin{equation} \phi_{c}^{(1)}(r)= \phi_{F}^{(1)}(r)+\phi_{\Pi}^{(1)}(r),\label{ads.41}\end{equation}
where the angles $\phi_{F}^{(1)}(r)$ and $\phi_{\Pi}^{(1)}(r)$ are
given in terms of the elliptic Jacobi integrals of the first and
third kind,
\begin{equation}
\phi_{F}^{(1)}(r)= \alpha_{1} \beta_{1} F(\psi_{c};\kappa_{c}),\qquad
\phi_{\Pi}^{(1)}(r)= \alpha_1 \gamma_{1} \Pi (\psi_{c};\kappa_{c},n_{c}),\label{ads42}
\end{equation}
respectively, where their parameters are
\begin{equation} \psi_{c}^{(1)}(r)=\arcsin
\sqrt{\frac{(r_{4}-r_{6})(r_{1}-r)}{(r_{1}-r_{4})(r-r_{6})}},\label{ads.43}\end{equation}
\begin{equation}
\kappa_{c}^{(1)}=\sqrt{\frac{(r_{1}-r_{4})(r_{5}-r_{6})}{(r_{1}-r_{5})(r_{4}-r_{6})}},\label{ads.44}\end{equation}
and
\begin{equation}
n_{c}^{(1)}=\frac{(r_{6}-r_{c})(r_{4}-r_{1})}{(r_{1}-r_{c})(r_{4}-r_{6})}.\label{ads.45}\end{equation}

\begin{figure}[!h]
 \begin{center}
   \includegraphics[width=175mm]{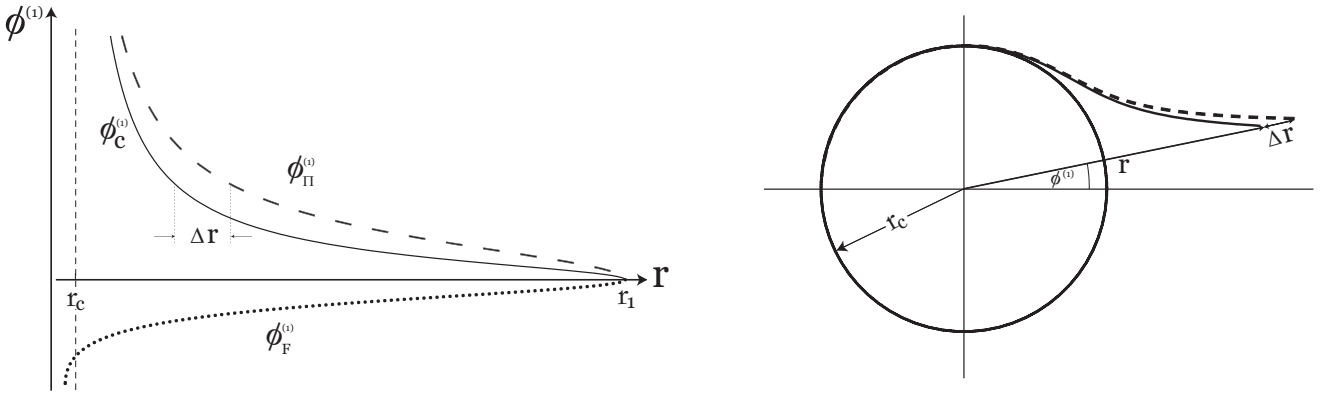}
 \end{center}
 \caption{Plot of the polar angle, $\phi_{c}^{(1)}$, as a function of the radial coordinate, $r$, with
 $M=1$, $Q=0.85M$, $\ell=10M$, $L=9M^{2}$, $q=0.18M$ and $m=0.2M$}
 \label{fig.4}
\end{figure}

In FIG. \ref{fig.4} we plot the polar angle, $\phi_{c}^{(1)}$, as
a function of the radial coordinate, $r$. We can  see that
$\phi_{\Pi}^{(1)}$ dominates the behavior of $\phi_{c}^{(1)}$
along the trajectory, specially when $r\rightarrow r_c$, in which
case $\phi_{F}^{(1)}$ takes a finite value.
This trajectory is characterized by an asymptotic tendency to the
unstable circle orbit from  distance bigger than the circle orbit
radius \textit{i.e.} in a infinite time we can found the particle
in such orbit.

 \subsubsection{Second class critical trajectory}
\begin{figure}[!h]
 \begin{center}
   \includegraphics[width=150mm]{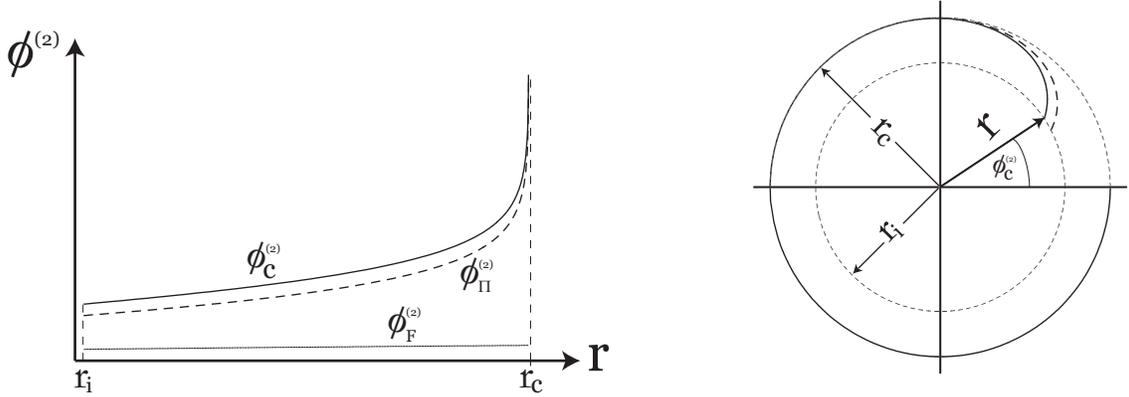}
 \end{center}
 \caption{Second class critical trajectory}
 \label{fig.5}
\end{figure}

In this case particles asymptotically tend to the circle orbit
from a distance less than the critical and the motion is realized
in the following range
$<r_{4}<r<r_{c}$. Then, we start
considering $\alpha_{1}$ (as was defined in previous case), the
constant $\beta_{2}=\frac{1 }{(r_{c}-r_{5})}$ and
$\gamma_{2}=\frac{ (r_{4}-r_{5})}{(r_{c}-r{4})(r_{c}-r_{5})}$,
therefore our solution is written as
\begin{equation} \phi_{c}^{(2)}(r)=\alpha_{1} \left[\beta_{2}
F(\psi_{c};\kappa_{c})+\gamma_{2} \Pi (\psi_{c};\kappa_{c},n_{c})
\right],\label{ads.46}\end{equation} where
\begin{equation} \psi_{c}^{(2)}=\arcsin
\sqrt{\frac{(r_{1}-r_{5})(r-r_{4})}{(r_{1}-r_{4})(r-r_{5})}},\label{ads.47}\end{equation}
\begin{equation}
\kappa_{c}^{(2)}=\sqrt{\frac{(r_{1}-r_{4})(r_{5}-r_{6})}{(r_{1}-r_{5})(r_{4}-r_{6})}},\label{ads.48}\end{equation}
and
\begin{equation}
n_{c}^{(2)}=\frac{(r_{5}-r_{c})(r_{1}-r_{4})}{(r_{4}-r_{c})(r_{1}-r_{5})},\label{ads.49}\end{equation}
As in the  last case the motion asymptotically tends to the unstable circular
orbit. For example in FIG. \ref{fig.5} we show the possible motion starting from a distance less than
the one corresponding to the unstable circular orbit and tending to it.

\section{Radial Trajectories} 
\begin{figure}[!h]
 \begin{center}
   \includegraphics[width=75mm]{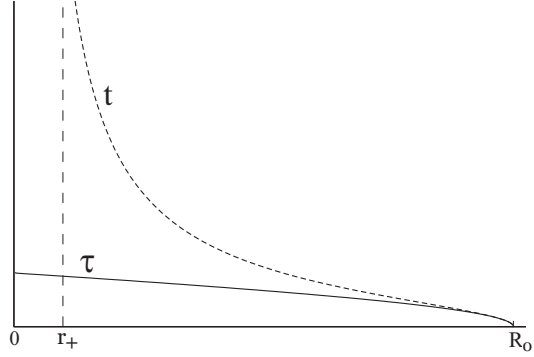}
 \end{center}
 \caption{Radial motion of the test particle for the proper time, $\tau$ (solid curve), and the
 coordinate time, $t$ (dashed curve). These solutions are very similar to
Schwarzschild case where in the proper time framework particles
can cross the event horizon and in the coordinate time framework,
event horizon acts as an asymptotic line. Both curves are plotted
by using $m=0.05 M$,  $Q=0.85 M$, $q=0.01 M$, $\ell=23 M$,
$E=0.067 M$}
 \label{fig.6}
\end{figure}

Radial motion corresponds to
a trajectory with vanished angular momentum. First at all, we are
considering the case of background of RNAdS, where $Q<M$. Let us
start with the relation
\begin{equation}
\frac{1}{f(r)}=\ell^{2}\frac{r^{2}}{F(r)},\label{ads.50}\end{equation}
where the fourth degree polynomial is given by
$F(r)=r^{4}+\ell^{2}(r^{2}-2Mr+Q^{2})=(r-r_{+})(r-r_{-})j(r)$. Now
the dynamical behavior is governed by
\begin{equation} \left(\frac{dr}{d\tau}\right)^{2} =
\left(E-\frac{qQ}{r}\right)^{2}- m^{2}\left(
1-\frac{2M}{r}+\frac{Q^{2}}{r^{2}}+\frac{r^{2}}{\ell^{2}}\right),\label{ads.51}\end{equation}
then, the trajectory equation can be written in the following form
\begin{equation} \left(\frac{dr}{d\tau}\right)^{2} =
\frac{m^{2}}{\ell^{2}}\frac{G(r)}{r^{2}},\label{ads.52}\end{equation}
where we  define the fourth degree polynomial $G(r)\equiv
\sum_{j=0}^{4} \alpha_{j} r^{j}$, with the coefficients
$\alpha_0=- Q^{2} \ell^{2}(1-q^{2}/m^{2})$, $\alpha_{1}=2
\ell^{2}(M-qQE/m^{2})$, $\alpha_{2}=\ell^{2}(E^{2}/m^{2}-1)$,
$\alpha_3=0$ and $\alpha_4 =-1$. The proper time quadrature is
\begin{equation} \tau(r)=\pm \frac{\ell}{m} \int
\frac{r}{\sqrt{G(r)}}dr,\label{ads.53}\end{equation} this motion
corresponds to the radial fall starting at rest from the distance $R_{0}$.
Then, considering negative sign,  the radial function in this case
is written as
$G(r)=(R_{0}-r)(r-\rho_{0})(r-\sigma_{1})(r-\sigma_{2})$, whose
four real roots are $ R_{0}> r> \rho_{0}>0
> \sigma_{1} > \sigma_{2}$. Then we can write our solution as
follows
\begin{equation}
 \tau(r)=k_{0}\Pi(\Psi;\kappa,\nu)+
k_{I}F(\Psi;\kappa),\label{ads.54}\end{equation} where
\begin{equation} k_{0}= \frac{\ell}{m} \frac{2(R_{0}-\sigma_{2})
}{\sqrt{(R_{0}-\sigma_{1})(\rho_{0}-\sigma_{2})}},\label{ads.55}\end{equation}

\begin{equation} k_{I}=\frac{\ell}{m} \frac{2\sigma_{2}
}{\sqrt{(R_{0}-\sigma_{1})(\rho_{0}-\sigma_{2})}}.\label{ads.56}\end{equation}

\begin{equation} \Psi= \arcsin
\sqrt{\frac{(\rho_{0}-\sigma_{2})(R_{0}-r)}{(R_{0}-\rho_{0})(r-\sigma_{2})}},\label{ads.57}\end{equation}

\begin{equation}
\kappa=\sqrt{\frac{(R_{0}-\rho_{0})(\sigma_{1}-\sigma_{2})}{(R_{0}-
\sigma_{1})(\rho_{0}-\sigma_{2})}},\label{ads.58}\end{equation}
and

\begin{equation}
\nu=\frac{\rho_{0}-R_{0}}{\rho_{0}-\sigma_{2}},\label{ads.59}\end{equation}
are constants. On the other hand, the corresponding coordinate
time quadrature is given by
\begin{equation} t(r)= \frac{\ell^{3}}{m} \int_{R_{0}}^{r}
\left(\frac{Er^{3}-qQr^{2}}{F(r)}\right)\frac{-dr}{\sqrt{G(r)}},\label{ads.60}\end{equation}
in order to see physical contents, it is necessary to separate in
partial fractions the  term under  the integral,

\begin{equation} \frac{Er^{3}-qQr^{2}}{F(r)}=
\frac{A}{r-r_{+}}+\frac{B}{r-r_{-}}+\frac{C+Dr}{j(r)},\label{ads.61}\end{equation}

\noindent where function   $j(r)=r^{2}+ar+b^{2}$, and the constant are given
by $a=r_{-}+r_{+}$ y $ b^{2}=\ell^{2}+a^{2}-r_{-}r_{+}$. One
important point is the physical behavior of the coordinate time
near the event horizon. There, the leading term is of the first
order, where the constant is $A=
\frac{r_{+}^{2}(Er_{+}-qQ)}{(r_{+}-r_{-})j(r_{+})}$. Therefore we
can write the general solution in this case as
\begin{equation} t(r)=
A_{0}t_{D}(r)+A_{1}t_{1}(r)+A_{2}t_{2}(r)+A_{3}t_{3}(r).\label{ads.62}\end{equation}
There, we can see that near horizon the coordinate term is given
by the leading term
\begin{equation} t(r)\approx A_{0}t_{D}(r),\label{ads.63}\end{equation}
where the constant is
\begin{equation}
A_{0}=\frac{\ell^{3}}{m}\frac{2A}{(R_{0}-r_{+})(r_{+}-\sigma_{2})\sqrt{(R_{0}-\sigma_{1})(\rho_{0}-\sigma_{2})}}
 ,\label{ads.64}\end{equation}
the exact solution for this term at the event horizons is
\begin{equation} t_{D}(r)= (R_{0}-\sigma_{2})\Pi (\Psi;\kappa,n)
-(R_{0}-r_{+})F(\Psi;\kappa),\label{ads.65}\end{equation} where
this solution is defined with the same set of parameters than the
proper time solution, but just we are added the constant
\begin{equation}n = \nu
\frac{r_{+}-\sigma_{2}}{r_{+}-R_{0}},\label{ads.66}\end{equation}
FIG. \ref{fig.6}, shows the behavior for the proper and coordinate time.
In simples terms, these solutions are very similar to
Schwarzschild case where in the proper time framework particles
can cross the event horizon and for the coordinate time framework,
event horizon acts as an asymptotic line.

%%%%%%%%%%%%%%%%%%%%%%%%%%%%%%%%%%%%%%%%%%%%%%%%%%%%%%%%%%%%%%%%%%%%%%%%%%%%%%%%%%%%%%%%

\section{ Motion of charged particles on the  Reissner-Nordstr\"{o}m de Sitter Black Hole }

In a similar way than the previous case of the  RNAdS black hole, we start with
particles with angular momentum $L\neq 0$, then radial quadrature
is given by
 \begin{equation}
 \left(\frac{dr}{d\tau}\right)^{2}=\left[E-V_{-}(r)\right]\left[E-V(r)\right],
 \label{ds.67}\end{equation}
then we can write the basic equation of the trajectories as follows \textbf{($\Lambda/3\equiv 1/\lambda^{2}>0$)}
\begin{equation}
\dot{r}^{2}=\left(\frac{m}{\lambda}\right)^{2}\frac{P_{6}(r)}{r^{4}},\label{ds.68}\end{equation}
where the characteristic polynomial for the allowed trajectories is
\begin{equation}
P_{6}(r)=\sum_{j=0}^{6} \omega_{j}
r^{j},\label{ds.69}\end{equation} the coefficients are given by
$$
\omega_0=-\frac{\lambda^{2} Q^{2} L^{2}}{m^{2}},\qquad \omega_1=\frac{2 \lambda^{2} M L^{2}}{m^{2}},\qquad
\omega_2=-\lambda^2\left(\frac{L^2}{m^2}+Q^2-\frac{q^2 Q^2}{m^2}\right),
$$
$$
\omega_3=2\lambda^2\left(M-\frac{q Q E}{m^2}\right),\qquad
\omega_4=\lambda^2\left(\frac{E^2}{m^2}-1+\frac{L^2}{m^2 \lambda^2}\right),\qquad
\omega_5=0,\qquad\omega_6=1.
$$

Using the constant $\zeta=\frac{\lambda L}{m}$, we can write the
angular quadrature as
\begin{equation} \phi(r)=\pm \zeta
\int^{r}\frac{dr}{\sqrt{P_{6}(r)}},\label{ds.70}\end{equation}

Now we  will give a detailed description of trajectories as in the
previous case. First we need to define the kind of motion in terms
of his energy and angular momentum against the effective
potential.

\begin{itemize}
\item Orbit with angular momentum: here we have, for a specific
combination of energy and angular momentum, the possibility to have
planetary orbits, when the effective potential exhibits a minimum.
Besides, we have three different kinds of orbits that depend on
the energy of the particle, compared to the effective potential.
The first kind orbits take place between the two maxima (that is planetary orbits).
The second kind orbits have the same energy than those from the first kind, but they fall to the event horizon, $r_+$.
Third kind orbits are those with the same energy than the cases before,
but they instead fall to the cosmologic horizon. $r_{++}$.

\item Radial Orbit ($L=0$) this orbit corresponds to a free fall to
cosmological horizon or event horizon.

\end{itemize}

\begin{figure}[!h]
 \begin{center}
   \includegraphics[width=125mm]{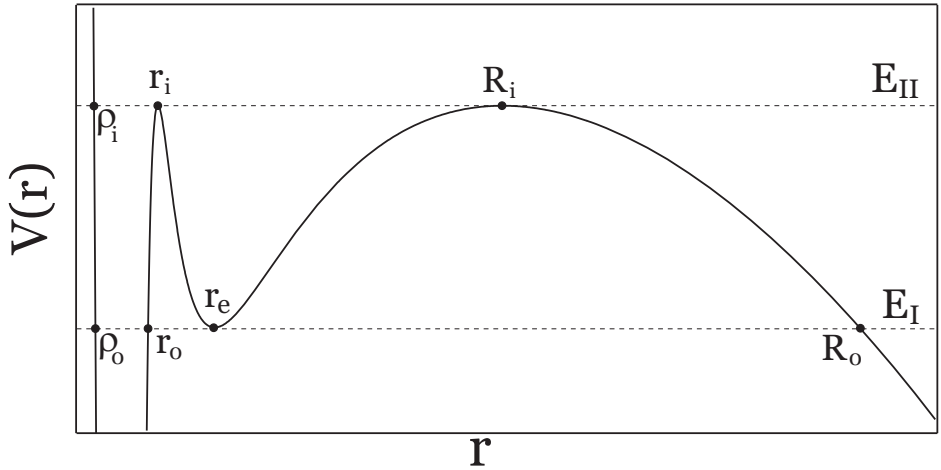}
 \end{center}
 \caption{Effective potential for charged particles on the geometry of RNdS for $M=1$, $Q=0.85M$, $\ell=400M$, $L=0.33M^{2}$, $q=0.18M$ and $m=0.2M$. }
 \label{fig7}
\end{figure}

\subsection{Second Kind trajectory: Fall to event horizon} In the
de Sitter case there are  trajectories that start from a finite
distance, at rest, and fall to the event horizon. The motion is
developed in the region $r_{0}\geq r \geq r_+
>\rho_{0}$. From (\ref{ds.70}) with the negative sign
and the characteristic polynomial  $P_{6}(r)=( R_{0}-r)(
r_{e}-r)^{2}( r_{0}-r)( r-\rho_{0})( r-\sigma_{0})$ we obtain the
integral motion as follows,

\begin{equation} \phi(r)=\zeta \int^{r}_{r_{0}}\frac{-dr}{( r_{e}-r)\sqrt{(
R_{0}-r)( r_{0}-r)( r-\rho_{0})(
r-\sigma_{0})}}.\label{a.14}\end{equation}

Therefore the solution
can be written as
\begin{equation} \frac{1}{\zeta
}\phi(r)=\alpha_{s}\left[\beta_{s}F(\varphi;k_{s})+\gamma_{s}\Pi(\varphi;k_{s},n_{s})\right],\label{a.15}\end{equation}
where the constants are given by
$\alpha_{s}=\frac{2}{\sqrt{(R_{0}-\rho_{0})(r_{0}-\sigma_{0})}}$,
$\beta_{s}=\frac{1}{(r_{e}-R_{0})}$ and
$\gamma_{s}=\frac{(R_{0}-r_{0})}{(R_{0}-r_{e})(r_{e}-r_{0})}$.
Also the parameters of the elliptic integrals are
 \begin{equation} k_{s}=\sqrt{\frac{(r_{0}-\rho_{0})(R_{0}-\sigma_{0})}{(R_{0}-\rho_{0})(r_{0}-\sigma_{0})}},\qquad
 n_{s}=\frac{(r_{0}-\rho_{0})(r_{e}-R_{0})}{(R_{0}-\rho_{0})(r_{e}-r_{0})},\quad \textrm{and}\quad
 \varphi(r)=\arcsin\sqrt{\frac{(R_{0}-\rho_{0})(r_{0}-r)}{(r_{0}-\rho_{0})(R_{0}-r)}},\label{d.21}\end{equation}

\begin{figure}[!h]
 \begin{center}
   \includegraphics[width=75mm]{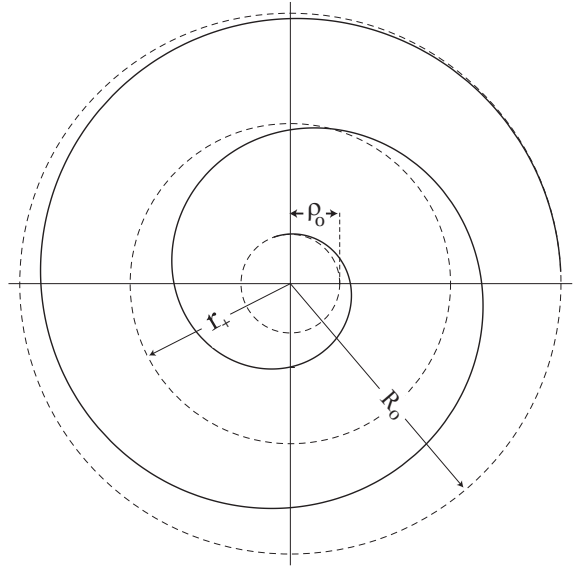}
 \end{center}
 \caption{Second Kind motion: It corresponds to trajectory starting
 from a distances bigger than event horizon  are doomed
 to cross it and fall to singularity.}
 \label{fig8}
\end{figure}

 Fig.(\ref{fig8}) shows us allowed trajectories of the charged particles
 in this case. Here, we can observe that the trajectories starting
 from a distances bigger than event horizon  are doomed
 to cross it and fall to singularity.

\subsection{Third Kind Trajectory: Fall to cosmological horizon}
This case corresponds to the motion that starts at a finite
distance from the event horizon and moves out from the field
forces, crossing the cosmological horizon. In this case the
allowed range is given by $r\geq R_{0}$. Integral Quadrature is
\begin{equation} \phi(r)=\zeta \int^{r}_{R_{0}}\frac{dr}{( r-r_{e})\sqrt{(
r-R_{0})( r-r_{0})( r-\rho_{0})(
r-\sigma_{0})}},\label{a.24}\end{equation} again the solution can
be written in terms of elliptical Jacobi integrals
\begin{equation} \frac{1}{\zeta
}\phi(r)=\alpha_{t}\left[\beta_{t}F(\varphi;k_{t})+\gamma_{t}\Pi(\varphi;k_{t},n_{t})\right],\label{a.25}\end{equation}
where the constants are given by
$\alpha_{t}=\frac{2}{\sqrt{(R_{0}-\rho_{0})(r_{0}-\sigma_{0})}}$,
$\beta_{t}=\frac{1}{(r_{0}-r_{e})}$ and
$\gamma_{t}=\frac{(R_{0}-r_{0})}{(R_{0}-r_{e})(r_{e}-r_{0})}$,
and the corresponding elliptic parameters are
 \begin{equation} k_{t}=\sqrt{\frac{(r_{0}-\rho_{0})(R_{0}-\sigma_{0})}{(R_{0}-\rho_{0})(r_{0}-\sigma_{0})}},\label{a.29}\end{equation}
 \begin{equation} n_{t}=\frac{(R_{0}-\sigma_{0})(r_{e}-r_{0})}{(r_{0}-\sigma_{0})(r_{e}-R_{0})},\label{a.30}\end{equation}
 \begin{equation} \psi(r)=\arcsin\sqrt{\frac{(r_{0}-\sigma_{0})(r-R_{0})}{(R_{0}-\sigma_{0})(r-r _{0})}},\label{a.31}\end{equation}

\begin{figure}[!h]
 \begin{center}
   \includegraphics[width=85mm]{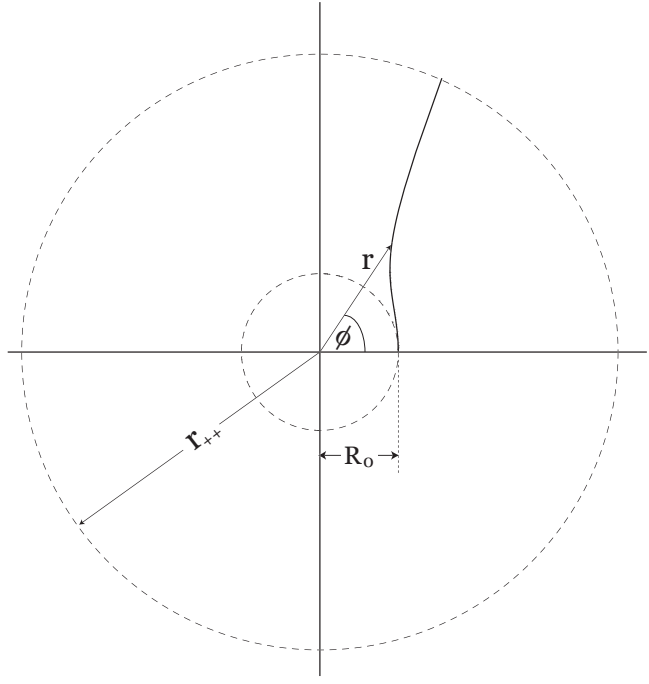}
 \end{center}
 \caption{Third Kind motion for a RNdS geometry: the particles moving out from the event horizon and then they are able to
cross the cosmological horizon losing casual connection with
physical observers.}
 \label{fig9}
\end{figure}

FIG.\ref{fig9} depicts   these  trajectories, that correspond to
particles moving out from the event horizon and then they are able to
cross the cosmological horizon losing casual connection with
physical observers.

\subsection{Critical Trajectory}
 As we have say, it is possible to obtain a class of orbits which posses two maxima at $r_i$ and $R_i$ ($r_i < R_i$)
 with the same value of the constant $E\equiv E_{II}$ (see FIG.\ref{fig7}). In this case, the conditions

\begin{equation}
 V(r_{i})=V(R_{i})\equiv E_{II},\label{ctd1}
\end{equation}
\noindent and
\begin{equation} \left(\frac{d V(r)}{dr}\right)_{r=R_{i}}=\left(\frac{d V(r)}{dr}\right)_{r=r_{i}}=0,\label{ctd2}
\end{equation}
\noindent  are satisfied. In this scenario, we can identify three possible trajectories: particles falling from an initial distance, $d_1$ ($d_1 < r_{++}$), to the unstable circular orbit at $R_i$ (first kind critical orbit); another one goes up to the unstable circular orbit at $r_i$ (second kind critical orbit) from an initial distance, $d_2$ ($d_2 > r_+$); and finally, the particles going to an asymptotically circular orbits at $R_i$ or $r_i$ (third kind critical orbit) depending on the sign of its initial velocity. Thus, the - (+) sign means that the particles falls (up) to the asymptotic orbit at $r_i$ ($R_i$).

\subsubsection{first kind critical orbit}
In this case, the particle starts its motion at $d_1$ and falls to the asymptotic orbit at $R_i$. From (\ref{ds.70}) with the negative sign
and the characteristic polynomial given by  $P_{6}(r)=(r -R_{0})^{2}(r-r_{i})^{2}( r-\rho_{i})( r-\sigma_{i})$, we obtain the solution

\begin{equation}
\phi_c^{(1)}(r)=\frac{\zeta}{\eta(R_{i}, r_{i})}\left[\frac{1}{\xi(R_i)}\ln\Theta(r, R_i)
-\frac{1}{\xi(r_i)}\ln\Theta(r, r_i) \right],
\label{ctd3}\end{equation}

\noindent where

\begin{equation}
\Theta(u, v)=2\xi(v)\left(\frac{\xi(u)+\xi(v)}{\eta(u, v)}\right)+\eta(v, \rho_i)+\eta(v, \sigma_i),
\label{ctd4}\end{equation}

\noindent with

\begin{equation}
\eta(u, v)= u - v,\quad \textrm{and}\quad
\xi(u)=\sqrt{\eta(u, \rho_i)\,\eta(u, \sigma_i)}.\label{ctd5}
\end{equation}

In left panel of FIG.\,\ref{fig10} we have shown a typical first kind trajectory.

\subsubsection{second kind critical orbit}

This orbits have trajectories that go up asymptotically to the unstable circular orbit at $r_i$ from a distance
$d_2$ ($d_2 > r_+$). Thus, we must choose the plus sign in eq. (\ref{ds.70}) with the characteristic polynomial given by
$P_{6}(r)=( R_{i}-r)^{2}( r_{i}-r)^{2}( r-\rho_{i})( r-\sigma_{i})$, in which case we obtain the solution

\begin{equation}
\phi_c^{(2)}(r)=-\frac{\zeta}{\eta(R_{i}, r_{i})}\left[\frac{1}{\xi(R_i)}\ln\left(-\Theta(r, R_i)\right)
-\frac{1}{\xi(r_i)}\ln\left(-\Theta(r, r_i)\right) \right],
\label{ctd6}\end{equation}

\noindent that we show in middle panel of FIG.\,\ref{fig10}.

\subsubsection{third kind critical orbit}

\begin{figure}[!h]
 \begin{center}
   \includegraphics[width=125mm]{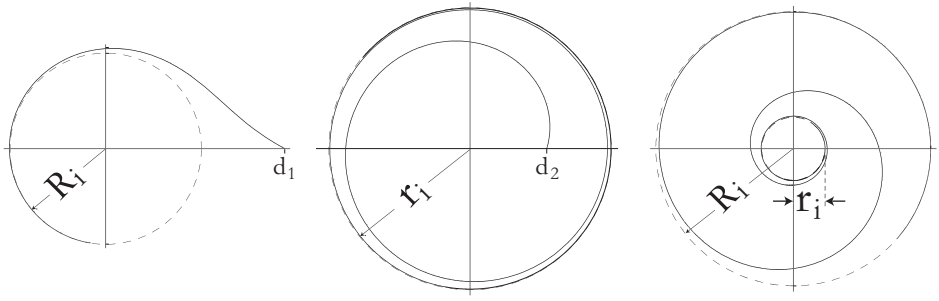}
 \end{center}
 \caption{Critical trajectories for charged particles in RNdS spacetime. Left panel: First kind orbit. The particle starts its motion from $d_1$
  ($\phi_{c}^{(1)}=0$) and approaches asymptotically to unstable circular orbit at $R_i$; Middle panel: Second kind orbit. The particle starts its motion from $d_2$  ($\phi_{c}^{(2)}=0$) and approaches asymptotically to unstable circular orbit at $r_i$. Right panel: Third kind orbit. The particle goes to the unstable circular orbit at $r_i$ or $R_i$, depending on the sign of its initial velocity; All graphics has been plotted using $M=1$, $m=0.1 M$, $Q=0.85 M$, $q=0.01 M$, $\lambda=400M$}
 \label{fig10}
\end{figure}

Finally, when the particle is at an initial distance $d_3$, such that $r_i \leq d_3 \leq R_i$, it has the ability to go asymptotically to the unstable circular orbit at $r_i$ or $R_i$, depending on the sign of its initial velocity. Therefore, the solution can be write as

\begin{equation}
\phi_c^{(3)}(r)=\pm\frac{\zeta}{\eta(R_{i}, r_{i})}\left[\frac{1}{\xi(R_i)}\ln\left(-\Theta(r, R_i)\right)
+\frac{1}{\xi(r_i)}\ln\Theta(r, r_i) \right]- \vartheta_0,
\label{ctd7}\end{equation}

\noindent where $\vartheta_0$ is a constant of integration that fixes the initial conditions and the - (+) sign means that the particle approaches
to the unstable circular orbit at $r_i$ ($R_i$). In right panel of FIG.\,\ref{fig10} we show this trajectories.

\section{Radial Motion }
The motion of radial particles ($L=0$) is described by the equation

\begin{equation} \left(\frac{dr}{d\tau}\right)^{2} =
\left(E-\frac{qQ}{r}\right)^{2}- m^{2}\left(
1-\frac{2M}{r}+\frac{Q^{2}}{r^{2}}-\frac{\Lambda
r^{2}}{3}\right).\label{rd.1}\end{equation}

The effective potential in this case is showed in

\begin{figure}[!h]
 \begin{center}
   \includegraphics[width=95mm]{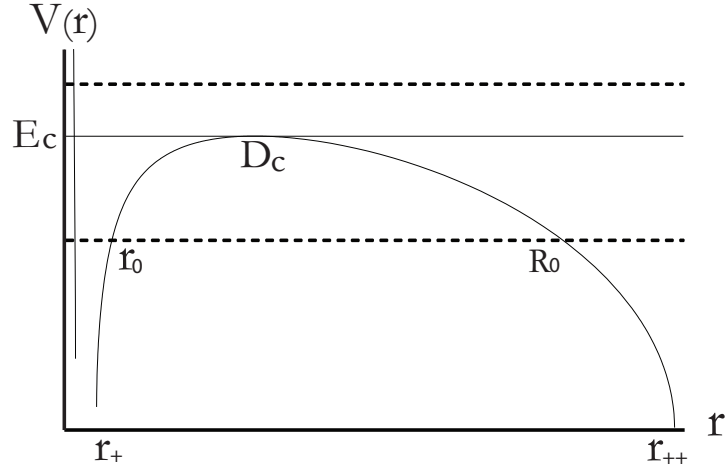}
 \end{center}
 \caption{Effective potential for radial charged particles in the RNdS spacetime  with $m=0.1$, $M=1$, $Q=0.8$, $q=0.01$, $\Lambda=400M$}
 \label{fig11}
\end{figure}

For different values of the constant of motion $E$, we have the following allowed orbits: The capture zone ($E<E_c$) and critical motion ($E=E_c$).

\subsection{Capture Zone} In this case, we have that $E<E_c$, and from here, we study two kind of orbits: first kind orbit, which fall
to the cosmological horizon, and second kind orbit, which fall to the event horizon. It is convenient to rewrite eq. (\ref{rd.1}) in the form

\begin{equation} \left(\frac{dr}{d\tau}\right)^{2} =
\frac{m^{2}}{\lambda^2}\frac{G(r)}{r^{2}},\label{rd.2}\end{equation}

\noindent where the
characteristic polynomial is
\begin{equation} G(r)=  r^{4}-\alpha r^{2}+\beta r-\gamma
,\label{ra.8}\end{equation} and the constants are given by
$\alpha=\lambda^2 (1-E^{2}/m^{2})$, $\beta=
2\lambda^2 (M-qQE/m^{2})$ and
$\gamma=\lambda^2 Q^{2}(1-q^{2}/m^{2})$.

Finally, we have
the following relations in order to obtain the motion integrals

\begin{equation}  \pm \frac{dr}{d\tau} =
\frac{m}{\lambda}\frac{\sqrt{G(r)}}{r},\label{rd.3}\end{equation}
\begin{equation}  \pm \frac{dt}{d\tau} =\lambda^2
\frac{r^{2}}{F(r)}\left(E-\frac{qQ}{r}\right),\label{rd.4}\end{equation}
\begin{equation}  \pm \frac{dr}{dt} =
\frac{m}{\lambda^{3}}F(r)\left(E-\frac{qQ}{r}\right)^{-1}\frac{\sqrt{G(r)}}{r^{3}}.\label{rd.5}\end{equation}

\subsubsection{First Kind: Fall to event horizon}
Radial motion  shows more variety depending on the roots of the
respective polynomial and the energy of the particles. This case
corresponds to particles that start from a finite distance bigger
than the last root of the effective potential and less than
cosmological horizon. Here, allowed region corresponds to $R_{0}
> r_{0} \geq r  > d_{1} > 0 > d_{2}$, such that the function $G(r)$ can be write as
$ G_{1}(r)=(r-R_{0})(r-r_{0})(r-d_{1})(r-d_{2})$, in which case, from eq. (\ref{rd.3}), we obtain the following solution for the proper time

\begin{equation}
 \tau^{(1)}(r)=\Omega_{1}\left[
F(\psi_{1};\kappa_{1})-\frac{R_0 - r_0}{R_0}\Pi
(\psi_{1};\kappa_{1},n_{1})\right],\label{rd.6}\end{equation}

\noindent where

$$ \Omega_{1}= \frac{\lambda}{m} \frac{2
R_{0}}{\sqrt{(R_{0}-d_{1})(r_{0}-d_{2})}},\quad \psi_{1}= \arcsin
\sqrt{\frac{(r_{0}-r)}{n_1 (R_{0}-r)}},\quad
\kappa_{1}=\sqrt{n_1 \frac{ (R_{0}-d_{2})}{(r_{0}-d_{2})}},\quad \textrm{and}\quad
n_{1}=\frac{r_{0}-d_{1}}{R_{0}-d_{1}}.
$$

On the other hand, in the coordinate time framework the general solution corresponds to
\begin{equation} t^{(1)}(r)=
A_{1}t_{I}(r)+A_{2}t_{II}(r)+A_{3}t_{III}(r)+A_{4}t_{IV}(r),\label{rd.7}\end{equation}

\noindent the divergent solution at the event horizon is $t_{II}$, which is given by

\begin{equation} t_{II}(r)= (r_{0}-r_{+})
F(\psi_{1};\kappa_{1})+(R_{0}-r_{0})\Pi
(\psi_{1};\kappa_{1},n_{2}),\label{rd.8}\end{equation}

\noindent with

$$
A_{2}=\frac{54}{m \lambda^{6}}\frac{1}{(R_{0}-r_{+})(r_{0}-r_{+})\sqrt{(R_{0}-d_{1})(r_{0}-d_{2})}},\quad \textrm{and}\quad
n_{2}=\sqrt{\frac{R_{0}-r_{+}}{r_{0}-r_{+}}\frac{r_{0}-d_{1}}{R_{0}-d_{1}}}.
$$

FIG. \ref{fig12} summarizes our result in this case and shows the fall
to the event horizon.

\subsubsection{Second Kind: Fall to cosmological horizon}

 Now, we are considering the radial motion of particles that fall to the
cosmological horizon. In this case, the radial coordinate belong to the interval $r_{++}>r \geq R_{0}$. So that, the solution
for the proper time is

\begin{equation}
 \tau^{(2)}(r)=\Omega_{2}\left[
F(\psi_{2};\kappa_{1})+\frac{R_0 - r_0}{r_0}\Pi
(\psi_{2};\kappa_{1},n_{3})\right],\label{rd.9}\end{equation}

\noindent where

$$ \Omega_{2}= \frac{\lambda}{m} \frac{2
r_{0}}{\sqrt{(R_{0}-d_{1})(r_{0}-d_{2})}},\quad \psi_{2}= \arcsin
\sqrt{\frac{(r-R_{0})}{n_3 (r-r_{0})}},\quad \textrm{and}\quad
n_{3}=\frac{R_{0}-d_{2}}{r_{0}-d_{2}}.
$$

In the coordinate time framework, after write the motion equation in terms of partial fractions, we obtain the general solution

\begin{equation} t^{(2)}(r)=
A_{1}t_{I}(r)+A_{2}t_{II}(r)+A_{3}t_{III}(r)+A_{4}t_{IV}(r),\label{rd.10}\end{equation}

\noindent the divergent solution at the cosmological horizon is $t_{I}$, which is given by

\begin{equation} t_{I}(r)= (r_{++}-R_{0})
F(\psi_{2};\kappa_{1})+(R_{0}-r_{0})\Pi
(\psi_{2};\kappa_{1},n_{4}),\label{rd.11}\end{equation}

\noindent with

$$
A_{1}=\frac{54}{m \lambda^{6}}\frac{1}{(r_{++}-R_{0})(r_{++}-r_{0})\sqrt{(R_{0}-d_{1})(r_{0}-d_{2})}},\quad \textrm{and}\quad
n_{4}=n_3 \frac{r_{++}-r_{0}}{r_{++}-R_{0}}.
$$

\begin{figure}[!h]
 \begin{center}
   \includegraphics[width=75mm]{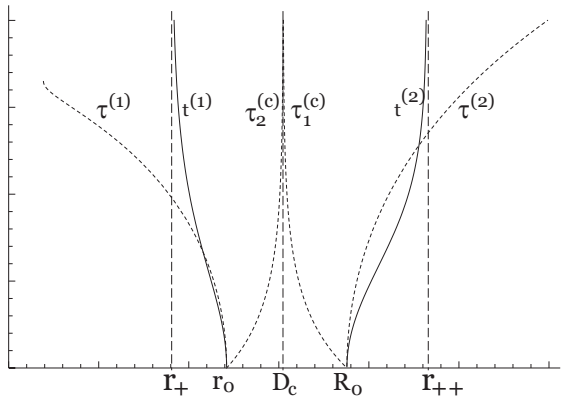}
 \end{center}
 \caption{Radial motion for charged particles in the coordinate time, $t$, and proper time, $\tau$ with $m=0.1$, $M=1$, $Q=0.8$, $q=0.01$, $\Lambda=400M$}
 \label{fig12}
\end{figure}

FIG. \ref{fig12} shows the behavior of the radial fall to event horizon.

\subsection{Radial Critical Trajectory}
One important case appears when a maximum in the effective potential exits. First
at all, we consider that $E <1 $ and $q <1 $, then we have four
real roots, where the maximum occurs at $r=D_{C}$,
\begin{equation} G_{II}(r)= (r-D_{C})^{2}g(r),\label{rd.12}\end{equation}
where
\begin{equation} g(r)= (r-D_{1})(r-D_{2}),\label{rd.13}\end{equation}
and the quadrature is
\begin{equation} \tau(r)=\pm \frac{3}{\lambda^2} \int
\frac{r}{|r-D_{C}|\sqrt{g(r)}}dr,\label{rd.14}\end{equation}

Here, we can classify orbits in terms of the distance where
the trajectory starts respect to $D_C$. We start
with critical first kind,  they correspond to trajectories that
asymptotically move to $D_C$ from greater distances $R_{0}$,
and the region under consideration is given by $R_{0} >  r
> D_{C} > D_{1} > 0 > D_{2}$. Then choosing negative sign for a
fall from $R_{0}$, we obtain
\begin{equation} \tau(r)= - \frac{3}{\lambda^2} \int_{R_{0}}^{r}
\left[1+\frac{D_{C}}{(r-D_{C})}\right]\frac{dr}{\sqrt{g(r)}},\label{ra.37}\end{equation}
whose general solution is
\begin{equation} \tau(r)= \frac{3}{\lambda^2}[\tau_{I}(r)+\tau_{II}(r)] ,\label{ra.38}\end{equation}
where the first integral is a regular function
\begin{equation}\tau_{I}(r)=
\ln\left[\frac{D_{C}+R_{0}+\sqrt{g(R_{0})}}{D_{C}+r+\sqrt{g(r)}}\right],\label{ra.40}\end{equation}
second integral is divergent at $D_{C}$
\begin{equation}
\tau_{II}(r)=\frac{1}{\sqrt{2}}\ln\left[\frac{R_{0}-D_{C}}{r-D_{C}}\right]
-\frac{1}{\sqrt{2}}\ln\left[\frac{4R_{0}+2\sqrt{2}\sqrt{g(R_{0})}}{4r+2\sqrt{2}\sqrt{g(r)}}\right],\label{ra.42}\end{equation}
Then, the final solution is given by
\begin{equation}  \frac{\lambda^2}{3} \tau^{(c)}_{1}(r)=
\ln\left[\frac{D_{C}+R_{0}+\sqrt{g(R_{0})}}{D_{C}+r+\sqrt{g(r)}}\right]+\frac{1}{\sqrt{2}}\ln\left[\frac{R_{0}-D_{C}}{r-D_{C}}\right]
-\frac{1}{\sqrt{2}}\ln\left[\frac{4R_{0}+2\sqrt{2}\sqrt{g(R_{0})}}{4r+2\sqrt{2}\sqrt{g(r)}}\right]
,\label{ra.43}\end{equation}

%\textbf{bla bla bla...}

The other kind corresponds to critical second kind
trajectories where particles approximate to critical radios from distances less than
$D_C$ and take infinite time to reach that distance.

 If the particle starts at $r_{0}$
and we choose minus sign, the solution is  given by
\begin{equation}  \frac{\lambda^2}{3} \tau^{(c)}_{2}(r)=
\ln\left[\frac{D_{C}+r_{0}+\sqrt{g(r_{0})}}{D_{C}+r+\sqrt{g(r)}}\right]+\frac{1}{\sqrt{2}}\ln\left[\frac{D_{C}-r_{0}}{D_{C}-r}\right]
-\frac{1}{\sqrt{2}}\ln\left[\frac{4r_{0}+2\sqrt{2}\sqrt{g(r_{0})}}{4r+2\sqrt{2}\sqrt{g(r)}}\right]
,\label{ra.45}\end{equation}

\section{Final remarks}
In this paper we  studied in detail the motion  of charged
particles in the vicinity of the Reissner-Nordstr\"{o}m (Anti)-dex
Sitter black hole. We  found, in the AdS case that exist
different kinds of trajectories depending on the angular momentum
of the particles. Specifically we found that exist

\begin{itemize}

\item  Planetary Orbit: In this case the orbit corresponds to a
bounded trajectory that exhibit oscillation between two extremal
distance (periastron and apastron). Besides, this trajectory
have one particular solution called Reissner-Nordstr\"{o}m limit.

\item  Second Kind Trajectory: This trajectory
 is computed with the same parameter than planetary orbits. It
 corresponds to a trajectory that starts  at rest from a
 finite distance less than periastron. This kind of motion represents the fall to the event
horizon, and it is considered that have a turning point inside to
cauchy horizon. Also it is shown
  an approximated  solution called Reissner-Nordstr\"{o}m limit

\item  Critical Trajectories: There are trajectories of the first and
second kind. The first one, starts at rest from a finite distance
from the outside of the unstable circular orbit and then   approximates
asymptotically to it. The second kind  approximates to
the unstable orbit from a  distance at the inner side of it.

 \item  Radial Trajectory: This trajectory has vanishing angular momentum and physically describes a
radial fall  to the event horizon starting at rest.

\item  Circle Orbits: For some fixed values of the constants it is
possible to find solutions of the equation of motion that
represent stable
  and unstable circle orbits.
\end{itemize}
In the dS case we mainly find two branches of classification that
are summed as follow
\begin{itemize}
\item Orbit with angular momentum: here we have for a specific
combination of energy and angular momentum, the possibility to have
planetary orbits, when the effective potential shows a minimum.
Besides, we have three different kinds of orbits that depend on the
energy of the particle compared with effective potential. First
kind, corresponds to an orbit whose energy is less than the global maximum of the effective
potential and starts at a large distance (bigger than the last root
of the respective polynomial) and the particles are doomed to go
 to the cosmological horizon. Second kind, are represented by
orbits that start at a distance between the event horizon and the
first root of the polynomial, they are doomed to fall  to the
event horizon. In this case  the energy of the particle is
also less than the global maximum of the effective potential. Third kind, corresponds to orbits
whose energy is bigger than the effective potential and particles
have  possibilities to move to cosmological or event horizon depending on the initial conditions in the velocity.

\item Radial Orbit ($l=0$) this orbit correspond to free fall to
cosmological horizon or event horizon. Also, we can find
the three kinds described for the orbit with angular momentum
\end{itemize}
In sum we have characterized through a specialized study all the
orbits for the black holes under consideration. We left the study
of photons for a future work.
%%%%%%%%%%%%%%%%%%%%%%%%%%%%%%%%%%%%%%%%%%%%%%%%%%%%%%%%%%%%%
%%%%%%%%%%%%%%%%%%%%%%%%%%%%%5

\begin{acknowledgments}
 In
the initial stages of this work, we benefited particularly from
insights of N. Cruz. This work was supported by COMISION NACIONAL
DE CIENCIAS Y TECNOLOGIA through FONDECYT Grant 1090613 (JS). This
work was also partially supported by PUCV DII (JS). J. S. and M.O. wish to
thank Departamento de F\'{\i}sica of the Universidad de Tarapac\'{a}
of Arica for its kind hospitality. C.L. was supported by
Universidad de Tarapac\'{a} Grant 4722-09.

\end{acknowledgments}

\end{document}